\documentclass[aps,prb,twocolumn,superscriptaddress,nofootinbib]{revtex4-2}

\usepackage{amsmath}
\usepackage{amssymb}
\usepackage{bm}
\usepackage{braket}
\usepackage{graphicx}
\usepackage{tikz}
\usepackage[colorlinks=true,linkcolor=blue,citecolor=blue,urlcolor=blue]{hyperref}

\graphicspath{{figures/}}

\begin{document}

\title{Quantum-centric assessment of saturated ferromagnetism beyond the flat-band limit}

\author{Kosuke Nogaki}
\email[]{kosuke.nogaki.scphys@niigata-u.ac.jp}
\affiliation{Department of Physics, Niigata University, Niigata 950-2181, Japan}
\affiliation{Institute for Research Administration, Niigata University, Niigata 950-2181, Japan}
\affiliation{RIKEN Center for Emergent Matter Science (CEMS), Wako 351-0198, Japan}

\author{Tomonori Shirakawa}
\affiliation{RIKEN Center for Computational Science (R-CCS), Kobe 650-0047, Japan}
\affiliation{RIKEN Center for Quantum Computing (RQC), Wako 351-0198, Japan}
\affiliation{Computational Condensed Matter Physics Laboratory, RIKEN Pioneering Research Institute (PRI), Wako 351-0198, Japan}
\affiliation{RIKEN Center for Interdisciplinary Theoretical and Mathematical Sciences (iTHEMS), Wako 351-0198, Japan}

\author{Seiji Yunoki}
\affiliation{RIKEN Center for Emergent Matter Science (CEMS), Wako 351-0198, Japan}
\affiliation{RIKEN Center for Computational Science (R-CCS), Kobe 650-0047, Japan}
\affiliation{RIKEN Center for Quantum Computing (RQC), Wako 351-0198, Japan}
\affiliation{Computational Condensed Matter Physics Laboratory, RIKEN Pioneering Research Institute (PRI), Wako 351-0198, Japan}

\author{Ryotaro Arita}
\affiliation{RIKEN Center for Emergent Matter Science (CEMS), Wako 351-0198, Japan}
\affiliation{Department of Physics, University of Tokyo, Hongo, Tokyo 113-0033, Japan}

\date{\today}

\begin{abstract}
Saturated ferromagnetism has been proven for every on-site repulsion $U > 0$ in Hubbard models with a flat lowest band, and the interaction required for it grows with the band width $W$ when the band becomes dispersive.
We study the dispersive case at quarter filling by quantum-selected configuration interaction and sample-based quantum diagonalization with IBM Quantum's Heron processors, focusing on the quantized square of the total spin of the computed states.
We treat a sawtooth chain of 16 unit cells on 64 qubits, whose full configuration space of $1.1\times 10^{14}$ determinants exceeds that of one of the largest full configuration interaction calculations, and obtain the stability diagram of saturated ferromagnetism on the decorated square lattice in the plane of $U$ and $W$.
For the chain, a subspace of only $3\times 10^{-5}$ of the full configuration space gives a state with more than $97\%$ of its weight in the sector of maximal spin.
For the decorated square lattice of $4\times 3$ unit cells on 48 qubits, the computed states in subspaces of less than $1\%$ of the full configuration space switch from low-spin states to nearly fully polarized states at $U \simeq 2W$.
The calculations also give a rigorous lower bound of $1.9W$ on the critical interaction of the $4\times 3$ lattice.
\end{abstract}

\maketitle

\section{Introduction}
\label{sec:intro}

Quantum many-body systems exhibit a variety of phenomena that cannot be understood within a single-particle picture~\cite{Fazekas1999}.
In strongly correlated electron systems, the competition between the kinetic energy of electrons and their Coulomb repulsion gives rise to unconventional superconductivity~\cite{SigristUeda1991,LeeNagaosaWen2006,Scalapino2012,Yanase2003}, magnetism~\cite{Yosida1996,Auerbach1994}, multipole ordering~\cite{WatanabeYanase2018,Hayami2018,Yatsushiro2021,HayamiKusunose2024}, and topological order~\cite{Wen2007,Fradkin2013}.
Which of these states is realized depends on the lattice, the band structure, and the strength of the interaction, and a small change of these parameters can turn one ground state into another~\cite{AritaOnodaKurokiAoki2000,AritaKurokiAoki2000,Georges1996,Kotliar2006}.
Understanding such phenomena from microscopic models is a central goal of condensed matter physics, and it requires an accurate description of the ground state.

Exact diagonalization, called full configuration interaction (FCI) in quantum chemistry, represents the ground state in the space of all Slater determinants, whose dimension grows exponentially with the number of orbitals.
One of the largest FCI calculations treats $1.3\times 10^{12}$ determinants on a supercomputer~\cite{Gao2024}, which corresponds to a Hubbard model of only a few tens of sites.
Many of the phenomena above, however, involve spontaneous symmetry breaking, which is sharply defined only in the thermodynamic limit.
Their study requires lattices large enough to control finite-size effects, whereas exact calculations are restricted to much smaller lattices.

Quantum computers are expected to overcome this limitation, because $n$ qubits represent a state in a space of dimension $2^{n}$, as envisioned by Feynman~\cite{Feynman1982}.
On fault-tolerant quantum computers, the quantum phase estimation algorithm~\cite{Kitaev1995,Cleve1998}, which uses the quantum Fourier transform~\cite{Coppersmith1994,Shor1997}, gives the eigenenergies of a many-body Hamiltonian~\cite{AbramsLloyd1999,AspuruGuzik2005}.
Its circuits, however, are too deep for current devices~\cite{NielsenChuang2010}.
For the pre-fault-tolerant devices with more than one hundred qubits that are available today, the variational quantum eigensolver (VQE) has been proposed~\cite{Peruzzo2014}.
The noise in the measured energies and the barren plateaus of the parameter landscape, however, make the variational optimization difficult~\cite{McClean2018,Cerezo2021,Tilly2022}.
As an alternative, quantum-classical hybrid methods have been developed in the framework of quantum-centric supercomputing, in which quantum processors and classical supercomputers are combined so that each performs the part of a calculation that suits it~\cite{Alexeev2024}.

In quantum-selected configuration interaction (QSCI) and sample-based quantum diagonalization (SQD), the configurations sampled from a quantum circuit span a subspace, in which the Hamiltonian is diagonalized~\cite{Kanno2023,RobledoMoreno2025}.
The resulting energy is a variational upper bound on the ground-state energy, because hardware noise affects only the choice of the configurations.
QSCI/SQD has been applied to molecules beyond the scale of exact diagonalization~\cite{RobledoMoreno2025}.
Since its proposal, QSCI/SQD has been extended, analyzed, and applied in a large number of studies~\cite{Nakagawa2024,Kaliakin2025a,Barison2025,Liepuoniute2025,Shajan2025,Sugisaki2025a,Mikkelsen2025,Nutzel2025,Reinholdt2025,Kaliakin2025b,Yoshida2025,Bauer2025,Danilov2025,Duriez2026,Barroca2026,Shirai2025,Ohgoe2025,Hafid2025,Nogaki2026,Chen2026,Erhart2025,Bazayeva2026a,Sriluckshmy2026,Smith2025,Weaving2025,Hossain2026,Khamadja2025,Shivpuje2026,Rocca2026,Sugisaki2025b,SBD2025,Hassman2025,Patra2026a,Raisuddin2025,Wray2025,Kanazawa2025,Patra2026b,Wang2025,Shajan2026,Ghasemi2026,Walkup2026,Doi2026,McFarthing2026,Yoshida2026,Aydeniz2026,Lotstedt2026,Park2026a,Graves2026,Miura2026,Anurag2026a,Kemmoku2026,Li2026,Gaberle2026,Stockinger2026,Lin2026a,Tran2026,Sabater2026,Zeng2026,Elgammal2026a,Xu2026,Do2026,Jones2026,Chang2026,Das2026a,Anurag2026b,Sosa2026,Park2026b,Zheng2026,David2026,Kaliakin2026,Bhuiyan2026,Patra2026c,Ligorio2026,Elgammal2026b,Sahoo2026,Das2026b,Bao2026,Park2026c,Vargas2026a,Kamoshita2026,VanRossum2026,Pandey2026,Vargas2026b,Agarwal2026,Lin2026b,Tajima2026,MuNoz2026,Tiwari2026,Fujimoto2026,Dulal2026,Kumar2026,Mostame2026,Iiyama2026,Patra2026d,Staros2026,Deschenes2026}.
In these applications, the main goal has been to obtain states with lower energies.
QSCI/SQD also yields an approximate wave function as an explicit vector in the subspace, from which physical quantities other than the energy can be evaluated.
For a quantity that is not quantized, however, it is difficult to estimate its exact value from an approximate wave function.

The exact value of a quantity with quantized eigenvalues is easier to estimate, because it is restricted to a discrete set of eigenvalues, with which an approximate expectation value can be compared.
In the special case where the eigenvalue of interest is the largest or smallest one, the variational principle also applies, and the deviation of the expectation value from this eigenvalue bounds the weight of the state outside the corresponding eigenspace from above.
In this work, we use the square of the total spin, $\bm{S}^2$, whose largest eigenvalue characterizes saturated ferromagnetism, that is, the fully polarized states, so that this special case applies.

Saturated ferromagnetism in the Hubbard model provides a problem in which both the total spin and the energy of the computed states can be related to exact statements.
Ferromagnetism of itinerant electrons is a long-standing problem, and rigorous results on it are rare~\cite{Tasaki1998,Tasaki2020}.
Nagaoka showed that the ground state is saturated ferromagnetic for $U = \infty$ with a single hole doped into the half-filled band on a class of lattices~\cite{Nagaoka1966}, and Lieb proved that the total spin of the ground state of the half-filled Hubbard model on a bipartite lattice is fixed by the difference between the numbers of sites on the two sublattices~\cite{Lieb1989}.
Mielke and Tasaki proved that the ground state is saturated ferromagnetic for every $U > 0$ in Hubbard models with a flat lowest band~\cite{Mielke1991L73,Mielke1991LineGraphs,Tasaki1992,MielkeTasaki1993}.
When the band becomes dispersive, saturated ferromagnetism survives for a sufficiently large $U$~\cite{Tasaki1995,Tasaki2003}, and its stability has been studied both rigorously~\cite{Tasaki1994,Tasaki1996,TanakaUeda2003} and numerically~\cite{LiangPang1995,Penc1996,SakamotoKubo1996,DaulNoack1997,Kohno1997,WatanabeMiyashita1997,DaulNoack1998,AritaAoki2000}.
Exact ferromagnetic ground states have also been constructed in a nearly flat band, where the ferromagnetism is stabilized by the quantum geometry of the band~\cite{Kitamura2026}.
The interaction required for saturated ferromagnetism grows with the band width, so that its value is a quantitative question that calls for accurate calculations on large lattices.

In this paper, we study saturated ferromagnetism in the Hubbard model on the sawtooth chain and the decorated square lattice with a dispersive lower band by QSCI/SQD with samples from IBM Quantum's Heron processors.
We find that focusing on the quantized square of the total spin allows us to draw physical conclusions on saturated ferromagnetism from subspaces that are a very small fraction of the full configuration space, only $3\times 10^{-5}$ of the full sector for the chain and less than $1\%$ for the $4\times 3$ lattice.
Combined with the exact energy of the fully polarized state, these calculations also give rigorous lower bounds on the critical interaction.

The rest of this paper is organized as follows.
In Sec.~\ref{sec:model}, we introduce the Hubbard model on the two lattices, construct the fully polarized reference state, and formulate the bound on the spin weight and the variational criterion.
In Sec.~\ref{sec:methods}, we describe the LUCJ circuits, the sampling on the quantum devices, the configuration recovery, and the subspace diagonalization.
The results for the chain, the $4\times 3$ lattice, and the stability diagram are presented in Secs.~\ref{sec:results:1d}, \ref{sec:results:4x3}, and \ref{sec:results:diagram}, respectively.
Section~\ref{sec:discussion} discusses the role of quantized quantities in QSCI/SQD, and Sec.~\ref{sec:summary} summarizes this paper.
Appendix~\ref{app:hardware} gives the qubit layouts and the characteristics of the quantum processors, and Appendix~\ref{app:raw} compares the results for the $4\times 3$ lattice with those of calculations without configuration recovery in the initial iteration.

\section{Model and spin diagnostics}
\label{sec:model}

\subsection{Lattices and Hamiltonian}
\label{sec:model:lattice}
\begin{figure}
  \centering
\begin{tikzpicture}[
  asite/.style={circle,draw,fill=black,inner sep=0pt,minimum size=5pt},
  bsite/.style={circle,draw,fill=white,inner sep=0pt,minimum size=5pt},
  hop/.style={line width=0.6pt},
  hop2/.style={line width=0.7pt,dash pattern=on 3pt off 1.5pt on 0.8pt off 1.5pt},
  hopab/.style={line width=0.7pt,dash pattern=on 0.8pt off 1.4pt},
  scale=0.75]
  \begin{scope}
    \node[anchor=south west] at (-0.4,1.0) {(a)};
    \foreach \i in {0,...,4} { \draw[hop] (\i,0) -- (\i+1,0); }
    \foreach \i in {0,...,4} {
      \draw[hopab] (\i,0) -- (\i+0.5,0.7) -- (\i+1,0);
      \node[bsite] at (\i+0.5,0.7) {};
    }
    \foreach \i in {0,...,5} { \node[asite] at (\i,0) {}; }
  \end{scope}
  \begin{scope}[xshift=7.0cm,yshift=-1.05cm,scale=1.2]
    \node[anchor=south west] at (-0.45,2.1) {(b)};
    \foreach \i in {0,1,2} { \draw[hop2] (\i,0) -- (\i,2); \draw[hop2] (0,\i) -- (2,\i); }
    \foreach \i in {0,1} { \foreach \j in {0,1} {
      \draw[hopab] (\i,\j) -- (\i+0.5,\j+0.5) -- (\i+1,\j+1);
      \draw[hopab] (\i+1,\j) -- (\i+0.5,\j+0.5) -- (\i,\j+1);
      \draw[hop] (\i,\j) to[bend left=22] (\i+1,\j+1);
      \draw[hop] (\i+1,\j) to[bend left=22] (\i,\j+1);
      \node[bsite] at (\i+0.5,\j+0.5) {};
    } }
    \foreach \i in {0,1,2} { \foreach \j in {0,1,2} { \node[asite] at (\i,\j) {}; } }
  \end{scope}
  \begin{scope}[yshift=-1.55cm, every node/.style={font=\footnotesize}]
    \node[asite] at (0.2,0) {}; \node[right] at (0.3,0) {$a$ site};
    \node[bsite] at (1.9,0) {}; \node[right] at (2.0,0) {$b$ site};
    \draw[hopab] (3.6,0) -- (4.3,0); \node[right] at (4.3,0) {$a$--$b$ hopping $\alpha t$};
    \draw[hop] (0.0,-0.6) -- (0.7,-0.6); \node[right] at (0.7,-0.6) {$a$--$a$ hopping $t$};
    \draw[hop2] (3.6,-0.6) -- (4.3,-0.6); \node[right] at (4.3,-0.6) {$a$--$a$ hopping $2t$};
  \end{scope}
\end{tikzpicture}
  \caption{Flat-band lattices on which the Hubbard model of Eq.~\eqref{eq:hubbard} is defined. (a) Sawtooth chain. The $a$ sites form a ring, and each $b$ site is placed above a bond of the ring and is connected to the two $a$ sites at the ends of that bond. (b) Decorated square lattice. The $a$ sites form a square lattice with nearest-neighbor and diagonal hoppings, and each $b$ site sits at a plaquette center and is connected to its four corner $a$ sites. The diagonal hoppings are drawn as arcs to keep them apart from the $a$--$b$ bonds. The legend gives the site types and the hopping amplitudes. The $b$ sites carry the on-site energy $\beta t$.}
  \label{fig:lattices}
\end{figure}
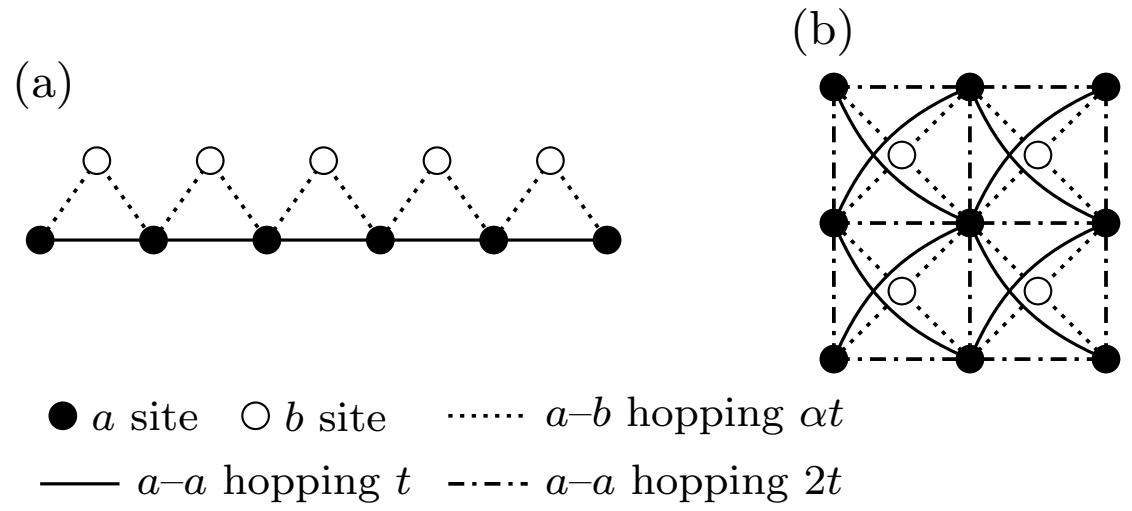

To study saturated ferromagnetism in nearly flat-band systems, we consider the Hubbard model
\begin{equation}
  H = H_{\mathrm{hop}} + U \sum_{i} n_{i\uparrow} n_{i\downarrow},
  \label{eq:hubbard}
\end{equation}
on two types of lattices introduced by Tasaki~\cite{Tasaki1992,MielkeTasaki1993,Tasaki1998,Tasaki2020}.
Here $c_{i\sigma}$ annihilates an electron of spin $\sigma$ on site $i$, $n_{i\sigma} = c^{\dagger}_{i\sigma} c_{i\sigma}$ counts the electrons of spin $\sigma$ on site $i$, and the positive constant $U$ is the on-site repulsion.
The hopping part $H_{\mathrm{hop}}$ is quadratic in $c_{i\sigma}$ and $c^{\dagger}_{i\sigma}$ and consists of the hoppings between neighboring sites and the on-site energies, which are specified below for each lattice.
Each lattice consists of $M$ unit cells, each of which contains one site of type $a$ and one site of type $b$, so that the number of sites is $L = 2M$.

One of the two lattices is the one-dimensional sawtooth chain shown in Fig.~\ref{fig:lattices}(a).
The $a$ sites form a chain of $M = N_x$ sites with nearest-neighbor hopping $t$.
Each $b$ site is placed above a bond of the chain and is connected to the two $a$ sites at the ends of that bond with hopping $\alpha t$.
The other is the two-dimensional decorated square lattice shown in Fig.~\ref{fig:lattices}(b), which is assembled from five-site cells consisting of one $b$ site and the four $a$ sites at the corners of a plaquette~\cite{Tasaki1998}.
The $a$ sites form an $N_x \times N_y$ square lattice of $M = N_x N_y$ sites with nearest-neighbor hopping $2t$ and diagonal hopping $t$, and each $b$ site is connected to its four corner $a$ sites with hopping $\alpha t$.
In both cases the $b$ sites carry the on-site energy $\beta t$, and we set $N_y = 1$ for the chain so that the expressions below apply to both lattices.

We take the lattice constant of the $a$ sites as the unit of length and assign to each $b$ site the position of the center of its bond or plaquette.
We write $a_{\bm{r}\sigma}$ and $b_{\bm{r}\sigma}$ for the operator $c_{i\sigma}$ on the $a$ or $b$ site at the position $\bm{r}$, so that $\bm{r}$ has integer coordinates for the $a$ sites and half-integer coordinates for the $b$ sites.
We impose periodic boundary conditions twisted by the angles $(\Phi_x, \Phi_y)$ and choose the gauge with a uniform vector potential, so that the hopping from the site at $\bm{r}$ to the site at $\bm{r}'$ carries the phase $e^{i\bm{\phi}\cdot(\bm{r}' - \bm{r})}$ with $\bm{\phi} = (\Phi_x/N_x, \Phi_y/N_y)$.

The hopping part of both lattices can be written in the form~\cite{Tasaki1992,MielkeTasaki1993}
\begin{equation}
  H_{\mathrm{hop}} = \alpha^{2} t \sum_{\bm{r},\sigma} B^{\dagger}_{\bm{r}\sigma} B_{\bm{r}\sigma}
  - z t N_e + (\beta - \alpha^{2} + z)\, t \sum_{\bm{r},\sigma} b^{\dagger}_{\bm{r}\sigma} b_{\bm{r}\sigma},
  \label{eq:hhop}
\end{equation}
where $\bm{r}$ runs over the positions of the $b$ sites and $N_e$ is the number of electrons.
The $a$ sites adjacent to the $b$ site at $\bm{r}$ are located at $\bm{r} + \bm{\delta}$ with $\bm{\delta} \in D$, where $D = \{\pm\bm{e}_x/2\}$ for the chain and $D = \{(\pm\bm{e}_x \pm \bm{e}_y)/2\}$ for the square lattice, and $\bm{e}_x$ and $\bm{e}_y$ are the unit vectors.
The number of elements of $D$ is denoted by $z$, which is $2$ for the chain and $4$ for the square lattice.
The operator $B_{\bm{r}\sigma}$ is defined by
\begin{equation}
  B_{\bm{r}\sigma} = b_{\bm{r}\sigma} + \frac{1}{\alpha} \sum_{\bm{\delta} \in D} e^{-i\bm{\phi}\cdot\bm{\delta}}\, a_{\bm{r}+\bm{\delta},\sigma}.
  \label{eq:bop}
\end{equation}
Expanding the first term of Eq.~\eqref{eq:hhop} generates the hoppings described above, with the amplitude of an $a$--$a$ hopping given by $t$ times the number of $b$ sites shared by the pair.
The expansion also produces the on-site energies $zt$ on every $a$ site and $\alpha^{2}t$ on every $b$ site, which the remaining two terms reset to zero and to $\beta t$.
When $\beta = \alpha^{2} - z$, the last term of Eq.~\eqref{eq:hhop} vanishes and the hopping Hamiltonian is a positive semidefinite operator shifted by $-ztN_e$.
The $M$ single-particle states annihilated by every $B_{\bm{r}\sigma}$ then form a flat lowest band at energy $-zt$, and we call $\beta = \alpha^{2} - z$ the flat-band condition.
Because the operators $B_{\bm{r}\sigma}$ of two $b$ sites adjacent to a common $a$ site are not orthogonal to each other, the remaining $M$ states do not form a flat band.
Under the flat-band condition, Tasaki proved that the ground state at quarter filling, $N_e = M$, is the fully polarized state for every $U > 0$~\cite{Tasaki1992,MielkeTasaki1993}.

Away from the flat-band condition the lowest band acquires a dispersion whose width is controlled by $\beta - \alpha^{2} + z$.
Throughout this work we set $\alpha = 2$ and measure energies in units of $t$, so that the flat-band condition reads $\beta = 0$ on the square lattice and $\beta = 2$ on the chain.
We vary $\beta$ to control the dispersion of the lowest band and fix the electron number to quarter filling, $N_e = M$, as in Tasaki's theorem.
The largest total spin at this filling is $S_{\max} = N_e/2$, and the single-particle dispersion is given in the next subsection.

\subsection{Band structure}
\label{sec:model:bands}

\begin{figure}
  \centering
  \includegraphics[width=\columnwidth]{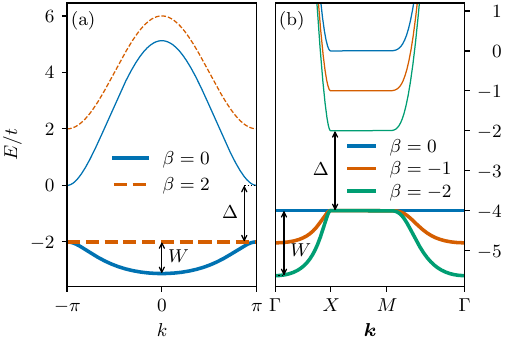}
  \caption{Single-particle band structure of Eq.~\eqref{eq:bands} for $\alpha = 2$ on the infinite lattice, where $\bm{q} = \bm{k}$. (a) Sawtooth chain at $\beta = 0$ (solid) and at $\beta = 2$ (dashed), which satisfies the flat-band condition. (b) Decorated square lattice along $\Gamma$--$X$--$M$--$\Gamma$ for $\beta = 0$, $-1$, and $-2$. Thick and thin lines are the lower band $E_-$ and the upper band $E_+$. The arrows mark the width $W$ of the lower band and the band gap $\Delta$ for the chain at $\beta = 0$ and for the square lattice at $\beta = -2$, respectively.}
  \label{fig:bands}
\end{figure}

Because the twist enters as a uniform vector potential, $H_{\mathrm{hop}}$ is block-diagonalized by the Fourier transforms $a_{\bm{k}\sigma} = M^{-1/2} \sum_{\bm{r}} e^{-i\bm{k}\cdot\bm{r}} a_{\bm{r}\sigma}$ and $b_{\bm{k}\sigma} = M^{-1/2} \sum_{\bm{r}} e^{-i\bm{k}\cdot\bm{r}} b_{\bm{r}\sigma}$, where $\bm{k}$ runs over the set $K$ of the $M$ wave vectors $(2\pi n_x/N_x, 2\pi n_y/N_y)$ with $n_x = 0, \ldots, N_x - 1$ and $n_y = 0, \ldots, N_y - 1$.
In the basis of $a_{\bm{k}\sigma}$ and $b_{\bm{k}\sigma}$, the $2\times 2$ block of $H_{\mathrm{hop}}$ depends on $\bm{k}$ only through the shifted wave vector $\bm{q} = \bm{k} - \bm{\phi}$ and is given by
\begin{equation}
  \mathcal{H}(\bm{q}) = t
  \begin{pmatrix}
    z\,[\,z\,g(\bm{q})^{2} - 1\,] & \alpha z\, g(\bm{q}) \\
    \alpha z\, g(\bm{q}) & \beta
  \end{pmatrix},
  \label{eq:hq}
\end{equation}
with $g(\bm{q}) = \cos(q_x/2)$ for the chain and $g(\bm{q}) = \cos(q_x/2)\cos(q_y/2)$ for the square lattice.
The eigenvalues of Eq.~\eqref{eq:hq} give the two bands
\begin{equation}
  E_{\pm}(\bm{q}) = \frac{t}{2} \Bigl[ \varepsilon(\bm{q}) + \beta
  \pm \sqrt{ \bigl( \varepsilon(\bm{q}) - \beta \bigr)^{2} + 4\alpha^{2} z^{2} g(\bm{q})^{2} } \,\Bigr]
  \label{eq:bands}
\end{equation}
with $\varepsilon(\bm{q}) = z\,[\,z\,g(\bm{q})^{2} - 1\,]$.
Under the flat-band condition $\beta = \alpha^{2} - z$ the lower band reduces to $E_-(\bm{q}) = -zt$ for every $\bm{q}$, in agreement with the construction in Sec.~\ref{sec:model:lattice}.
Figure~\ref{fig:bands} shows the band structure of the chain at $\beta = 0$ and $2$ and that of the square lattice at $\beta = 0$, $-1$, and $-2$.

On the infinite lattice, the two bands $E_\pm$ depend on $\bm{q}$ only through $g(\bm{q})^2$, which ranges from $0$ at the zone boundary to $1$ at $\bm{q} = \bm{0}$.
For $\beta < \alpha^{2} - z$, $E_-$ decreases and $E_+$ increases as $g^2$ grows from $0$ to $1$.
The top of the lower band and the bottom of the upper band therefore lie at $g = 0$, where $\mathcal{H}(\bm{q})$ is diagonal with the eigenvalues $-zt$ and $\beta t$.
For $\beta > -z$ the eigenvalue $-zt$ belongs to the lower band, so the top of the lower band stays at $-zt$ as $\beta$ varies and the band gap is $\Delta = (z + \beta)t$.
In contrast, the bottom of the lower band lies at $\bm{q} = \bm{0}$, where $g^{2} = 1$, and moves down as $\beta$ decreases.
We define the band width as $W = -zt - E_-(\bm{0})$.

The choice of the twist matters for detecting saturated ferromagnetism on a finite lattice.
Arita and Aoki showed that the fully polarized state is accompanied by a spin-singlet state with a spiral spin texture, in which the polarized electrons effectively feel the twist shifted by $\pi$ along the direction of the spiral, and that the spiral state can become the ground state of a finite system~\cite{AritaAoki2000}.
A spiral ground state has $\langle \bm{S}^2 \rangle = 0$ and cannot be identified as ferromagnetic from the total spin.
We therefore choose for each lattice the twist among $\Phi_x, \Phi_y \in \{0, \pi\}$ that minimizes the sum $\sum_{\bm{k} \in K} E_-(\bm{k} - \bm{\phi})$ of the $M$ levels of the lower band, so that shifting the twist by $\pi$ along either direction raises the energy of the filled lower band~\cite{AritaAoki2000}.
The resulting twists are $\Phi_x = \pi$ for the chain of 16 unit cells and $(\Phi_x, \Phi_y) = (\pi, 0)$ and $(\pi, \pi)$ for the $4\times 3$ and $4\times 2$ square lattices, that is, $\Phi = \pi$ along a direction with an even number of unit cells and $\Phi = 0$ along a direction with an odd number.

\subsection{Fully polarized reference state}
\label{sec:model:fm}

The reference state against which the computed states are compared is the fully polarized state.
Consider first a state $\ket{\psi}$ that contains only up-spin electrons.
Since $n_{i\downarrow}\ket{\psi} = 0$ on every site, the interaction term of Eq.~\eqref{eq:hubbard} acting on $\ket{\psi}$ gives zero and $H$ acts on $\ket{\psi}$ as $H_{\mathrm{hop}}$.
The eigenstates of $H$ with $N_e$ up-spin electrons are therefore Slater determinants of single-particle eigenstates of $H_{\mathrm{hop}}$, and their energies are sums of $N_e$ distinct single-particle energies that do not depend on $U$.
The eigenstates with only up-spin electrons have $S = S_z = S_{\max}$.

The Hamiltonian commutes with the total spin $\bm{S}$, because neither the hopping nor the twist depends on spin.
Each eigenstate of $H$ with $S = S_{\max}$ is therefore degenerate with the member of its multiplet with $S_z = S_{\max}$, which contains only up-spin electrons.
The lowest energy in the sector $S = S_{\max}$ is thus the sum of the $N_e$ lowest single-particle energies of $H_{\mathrm{hop}}$ for every $U$.

At quarter filling, $N_e = M$, and for $\Delta > 0$, the $N_e$ lowest single-particle levels are the $M$ levels of the lower band.
Let $d_{\bm{k}\sigma}$ annihilate an electron of spin $\sigma$ in the single-particle eigenstate of the lower band with wave vector $\bm{k}$, whose energy is $E_-(\bm{k} - \bm{\phi})$.
The lowest state with $S = S_z = S_{\max}$ is then the unique fully polarized state
\begin{equation}
  \ket{\mathrm{FM}, S_{\max}} = \prod_{\bm{k} \in K} d^{\dagger}_{\bm{k}\uparrow} \ket{0},
  \label{eq:fm}
\end{equation}
where $\ket{0}$ is the vacuum, and its energy is
\begin{equation}
  E_{\mathrm{FM}} = \sum_{\bm{k} \in K} E_-(\bm{k} - \bm{\phi}).
  \label{eq:efm}
\end{equation}
The twists chosen in Sec.~\ref{sec:model:bands} are those that minimize $E_{\mathrm{FM}}$.
Applying the total spin-lowering operator $S^{-}$ repeatedly to $\ket{\mathrm{FM}, S_{\max}}$ generates the $2S_{\max} + 1$ members $\ket{\mathrm{FM}, S_z} \propto (S^{-})^{S_{\max} - S_z} \ket{\mathrm{FM}, S_{\max}}$, which share the energy $E_{\mathrm{FM}}$.
The member $\ket{\mathrm{FM}, 0}$ has $S_z = 0$ and is the one that is directly compared with the states computed in Sec.~\ref{sec:methods}.
Saturated ferromagnetism in the ground state of $H$ is therefore equivalent to the statement that no state with $S < S_{\max}$ has an energy below $E_{\mathrm{FM}}$.

\subsection{Spin components and variational criterion}
\label{sec:model:criterion}

The states computed in Sec.~\ref{sec:results} are approximate, and we characterize them by two quantities that can be related to exact statements about $H$.
Consider a normalized state $\ket{\Psi}$ with $N_e$ electrons and decompose it as $\ket{\Psi} = \sum_{S} \sqrt{p_S}\, \ket{\Psi_S}$, where $\ket{\Psi_S}$ is a normalized state with total spin $S$ and the weights $p_S \geq 0$ satisfy $\sum_S p_S = 1$.
The eigenvalues of $\bm{S}^2$ are quantized as $S(S+1)$, but the expectation value
\begin{equation}
  \langle \bm{S}^2 \rangle = \sum_{S} p_S\, S(S+1)
  \label{eq:s2}
\end{equation}
takes any value between $0$ and $S_{\max}(S_{\max}+1)$.
The weight outside the sector of maximal spin, $\eta = 1 - p_{S_{\max}}$, is bounded by the deviation of $\langle \bm{S}^2 \rangle$ from its largest possible value.
Since $S$ changes in integer steps, every $S < S_{\max}$ satisfies $S(S+1) \leq S_{\max}(S_{\max}+1) - 2S_{\max}$, and Eq.~\eqref{eq:s2} gives
\begin{equation}
  \eta \leq \frac{S_{\max}(S_{\max}+1) - \langle \bm{S}^2 \rangle}{2 S_{\max}}.
  \label{eq:eps_bound}
\end{equation}
A value of $\langle \bm{S}^2 \rangle$ close to $S_{\max}(S_{\max}+1)$ thus guarantees that $\ket{\Psi}$ lies almost entirely in the sector $S = S_{\max}$.
The bound concerns the computed state itself and does not measure the overlap of $\ket{\Psi}$ with the ground state of $H$.

By the variational principle, the energy $E_{\Psi} = \bra{\Psi} H \ket{\Psi}$ is an upper bound on the ground-state energy $E_0$ of $H$.
We measure $E_{\Psi}$ from the energy of the fully polarized state as $\Delta E = E_{\Psi} - E_{\mathrm{FM}}$.
When $\Delta E < 0$, the ground-state energy satisfies $E_0 < E_{\mathrm{FM}}$.
Since every state with $S = S_{\max}$ has an energy of at least $E_{\mathrm{FM}}$, as shown in Sec.~\ref{sec:model:fm}, the ground state then has $S < S_{\max}$, and saturated ferromagnetism is excluded rigorously.
When $\Delta E \geq 0$, no conclusion about the ground state follows, because an approximate state can lie above $E_{\mathrm{FM}}$ even if the ground state has a low spin.

The criterion translates into a bound on the interaction strength at which saturated ferromagnetism sets in on a finite lattice.
For $U < U'$ the difference $H(U') - H(U) = (U' - U) \sum_i n_{i\uparrow} n_{i\downarrow}$ is positive semidefinite, so the ground-state energy $E_0(U)$ does not decrease with $U$.
Since $E_0(U) \leq E_{\mathrm{FM}}$ for every $U$ and $E_{\mathrm{FM}}$ does not depend on $U$, $E_0(U) = E_{\mathrm{FM}}$ at one value of $U$ implies the same equality at every larger $U$.
The fully polarized state is therefore a ground state exactly for $U \geq U_c$, where $U_c$ is the smallest $U$ with $E_0(U) = E_{\mathrm{FM}}$.
For a dispersive lower band $U_c$ is positive, since at $U = 0$ the state with the $M/2$ lowest levels of the lower band doubly occupied has an energy below $E_{\mathrm{FM}}$ for the even $M$ considered here.
Any $U$ at which a computed state has $\Delta E < 0$ satisfies $U < U_c$, and the largest such $U$ gives a rigorous lower bound on $U_c$.
An upper bound on $U_c$ does not follow from $\Delta E \geq 0$.

\section{Computational methods}
\label{sec:methods}

\subsection{Reference state, LUCJ circuit, and hardware sampling}
\label{sec:methods:circuit}

We employ the local unitary cluster Jastrow (LUCJ) ansatz~\cite{Matsuzawa2020,Motta2023} as the trial wave function on the quantum circuit:
\begin{equation}
  \ket{\Psi_{\mathrm{LUCJ}}} = e^{T_1 - T_1^{\dagger}} e^{K} e^{iJ} e^{-K} \ket{\Psi_{\mathrm{RHF}}},
  \label{eq:lucj}
\end{equation}
where $\ket{\Psi_{\mathrm{RHF}}}$ is the restricted Hartree--Fock (RHF) determinant.
Here, $T_1$ is the single-excitation operator of the coupled-cluster singles and doubles (CCSD) theory, and $f_{p\sigma}$ annihilates an electron of spin $\sigma$ in the $p$th RHF orbital.
The one-body operator $K = \sum_{pr,\sigma} K_{pr} f^{\dagger}_{p\sigma} f_{r\sigma}$ with an anti-Hermitian matrix $K_{pr}$ and the density-density operator $J = \sum_{pr,\sigma\tau} J_{p\sigma,r\tau} f^{\dagger}_{p\sigma} f_{p\sigma} f^{\dagger}_{r\tau} f_{r\tau}$ are obtained from the leading term of the double factorization of the CCSD double-excitation operator $T_2$~\cite{Matsuzawa2020,Motta2023}.
Following Ref.~\cite{RobledoMoreno2025}, $J$ contains all nearest-neighbor same-spin pairs $(p, p+1)$ and the opposite-spin pairs $(p, p)$ for which $p$ is a multiple of $4$, where the RHF orbitals are labeled by $p = 0, \ldots, 2M - 1$ in ascending order of their energies.
This choice is adapted to the heavy-hex coupling map of the processors, on which the qubits of the spin-up and spin-down orbitals are placed on two paths connected through single auxiliary qubits at $p = 0, 4, 8, \ldots$~\cite{RobledoMoreno2025}.
Since the paths are longer than the processor, this applies to $p \leq 16$, and the transpiler connects the remaining opposite-spin pairs through longer routes.

The RHF and CCSD calculations are performed with \texttt{PySCF}~\cite{PySCF2020} in the basis of $a_{\bm{k}\sigma}$ and $b_{\bm{k}\sigma}$ introduced in Sec.~\ref{sec:model:bands}.
The RHF determinant doubly occupies the $N_e/2$ lowest RHF orbitals and has $S = 0$.
The LUCJ circuits are generated with \texttt{ffsim}~\cite{ffsim2026} under the Jordan--Wigner mapping~\cite{JordanWigner1928}, which assigns the $4M$ spin orbitals to $4M$ qubits, $64$ for the chain with $M = 16$ and $48$ for the $4\times 3$ lattice.

The quantum experiments are run on IBM Quantum's \texttt{ibm\_kawasaki} and \texttt{ibm\_pittsburgh}, which are Heron processors equipped with 156 qubits.
The circuits are transpiled using \texttt{Qiskit}~\cite{Qiskit2024} with a fixed initial layout.
Each circuit is sampled in $10$ jobs of $10^{6}$ shots.
The qubit layout and the characteristics of the processors are given in Appendix~\ref{app:hardware}.

At large $U$ the CCSD iterations do not always converge, and we then diagonalize the Hamiltonian at the target $U$ with the samples of the circuit generated at the largest $U$ at which the CCSD calculation converges.
The computed energy remains an upper bound on the ground-state energy at the target $U$, because the samples enter only through the choice of the configurations.
In Sec.~\ref{sec:results}, the samples are taken from a circuit generated at a different value of $U$ only for the chain at $U = 5$.
The values of $U$ of the circuits used for the calculations of Appendix~\ref{app:raw} are listed in that appendix.

\subsection{Post-selection and configuration recovery}
\label{sec:methods:recovery}

Each measured bitstring represents, through the Jordan--Wigner mapping, a configuration of electrons in the up-spin and down-spin RHF orbitals.
Hardware noise produces configurations whose numbers of up-spin or down-spin electrons differ from $N_e/2$.
We correct these configurations by the self-consistent configuration recovery of Ref.~\cite{RobledoMoreno2025} as implemented in \texttt{qiskit-addon-sqd}~\cite{QiskitAddonSQD2024}.
For an up-spin or down-spin occupation string with a wrong number of electrons, the recovery flips the required number of bits, which are chosen stochastically with weights determined by the deviation of each bit from a reference occupation $\bar{n}_p$ of orbital $p$.
Occupation strings with the correct number of electrons are left unchanged.
We then post-select the configurations with $N_e/2$ electrons of each spin and randomly subsample them, weighted by their measured frequencies, into $n_b$ batches of $n_s$ configurations.

We perform five iterations, $i = 0, \ldots, 4$, of the self-consistent configuration recovery, each of which consists of the configuration recovery, the post-selection and subsampling, and the subspace diagonalization of Sec.~\ref{sec:methods:diag}.
The reference occupation $\bar{n}_p$ is updated as follows.
In iteration $i \geq 1$, $\bar{n}_p$ is the occupation of orbital $p$ in the states obtained by the diagonalization of Sec.~\ref{sec:methods:diag} in iteration $i-1$, averaged over the batches, and is the same for both spins.
In iteration $0$, we use one of two references.
The HF reference sets $\bar{n}_p = 1$ for the $N_e/2$ lowest RHF orbitals and $\bar{n}_p = 0$ otherwise, which is the occupation of the RHF determinant.
The FM reference sets $\bar{n}_p = 1/2$ for the $N_e$ lowest RHF orbitals and $\bar{n}_p = 0$ otherwise, so that each of these orbitals holds one electron on average.
We refer to the calculations started from the two references as HF-seeded and FM-seeded.

For the $4\times 3$ lattice, we use the FM reference with $n_b = 10$ and $n_s = 10^{4}$.
For the chain, we use both references with $n_b = 4$ and vary $n_s$ from $10^{3}$ to $6\times 10^{4}$ to control the dimension of the subspace.
Appendix~\ref{app:raw} compares the FM-seeded results of the $4\times 3$ lattice with those obtained without recovery in iteration $0$.

\subsection{Subspace diagonalization}
\label{sec:methods:diag}

For each batch, we collect the up-spin and down-spin occupation strings of its configurations into a single set $\mathcal{A}$ and add the string of the RHF determinant.
Following the closed-shell scheme of Ref.~\cite{RobledoMoreno2025}, the subspace is spanned by the determinants whose up-spin and down-spin strings both belong to $\mathcal{A}$, so that its dimension is $|\mathcal{A}|^{2}$.
Every determinant in the subspace has $N_e/2$ electrons of each spin.
The diagonalization is therefore performed in the sector $S_z = 0$, which contains the members with $S_z = 0$ of the multiplets with every $S$ from $0$ to $S_{\max}$.
The full sector $S_z = 0$ has the dimension $1.8\times 10^{10}$ for the $4\times 3$ lattice studied in Sec.~\ref{sec:results:4x3} and $1.1\times 10^{14}$ for the chain of 16 unit cells studied in Sec.~\ref{sec:results:1d}, and the latter exceeds the $1.3\times 10^{12}$ determinants of one of the largest full configuration interaction calculations~\cite{Gao2024}.

We approximate the lowest eigenstate of the Hamiltonian projected onto the subspace by the Davidson method implemented in the SBD library~\cite{SBD2025}, with the convergence threshold $10^{-4}$ for the $4\times 3$ lattice and $10^{-3}$ for the chain.
From the reduced density matrices of this state we obtain $\langle \bm{S}^2 \rangle$ and the orbital occupations used in Sec.~\ref{sec:methods:recovery}.
For each value of $U$, and of $n_s$ for the chain, we adopt the state with the lowest energy among the converged states of all iterations and batches, and report its energy and $\langle \bm{S}^2 \rangle$.

\section{Results}
\label{sec:results}

\subsection{Sawtooth chain at 64 qubits}
\label{sec:results:1d}

\begin{figure}[tb]
  \centering
  \includegraphics[width=\columnwidth]{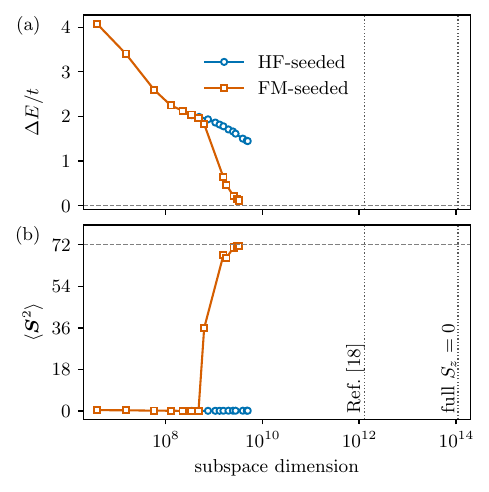}
  \caption{Energy and total spin of the computed states of the chain at $\beta = 0$ and $U = 5$ as functions of the subspace dimension $|\mathcal{A}|^{2}$.
    (a) Energy measured from $E_{\mathrm{FM}}$.
    (b) $\langle \bm{S}^2 \rangle$, where the dashed line marks its maximum $S_{\max}(S_{\max}+1) = 72$.
    Circles and squares show the HF-seeded and FM-seeded calculations, respectively.
    The dotted vertical lines mark the $1.3\times 10^{12}$ determinants of the full configuration interaction calculation of Ref.~\cite{Gao2024} and the dimension $1.1\times 10^{14}$ of the full sector $S_z = 0$.}
  \label{fig:1d}
\end{figure}

We first consider the chain with $M = 16$ unit cells, that is, $L = 32$ sites, at $\beta = 0$ and $U = 5$.
At $\beta = 0$ the chain is away from the flat-band condition $\beta = 2$, and the lower band has the width $W = (\sqrt{17} - 3)t \simeq 1.12t$ shown in Fig.~\ref{fig:bands}(a), so that $U/W \simeq 4.5$.
The quarter filling gives $N_e = 16$ and $S_{\max} = 8$, and the fully polarized state has the energy $E_{\mathrm{FM}} = -43.363t$.
As described in Sec.~\ref{sec:methods:circuit}, the samples are generated by the circuit at $U = 4.2$.
Figure~\ref{fig:1d} shows $\Delta E = E - E_{\mathrm{FM}}$ and $\langle \bm{S}^2 \rangle$ of the adopted states as functions of the subspace dimension, which we control by the number $n_s$ of configurations per batch.

The HF-seeded calculations give low-spin states at every dimension.
As the dimension increases from $3.5\times 10^{8}$ to $5.0\times 10^{9}$, the energy decreases smoothly and $\Delta E$ falls from $2.03t$ to $1.45t$, while $\langle \bm{S}^2 \rangle$ stays below $0.10$.
The energy alone thus shows a steady improvement of the computed state, and only $\langle \bm{S}^2 \rangle$ reveals that the state remains far from the fully polarized multiplet.

The FM-seeded calculations also give low-spin states up to $n_s = 1.2\times 10^{4}$, where the dimension is $4.8\times 10^{8}$ and the energy is close to that of the HF-seeded state of the same dimension.
Between $n_s = 1.2\times 10^{4}$ and $3.6\times 10^{4}$, $\langle \bm{S}^2 \rangle$ increases to $67.5$ and $\Delta E$ drops to $0.64t$.
At the largest dimension $3.3\times 10^{9}$, which is only $3\times 10^{-5}$ of the full sector $S_z = 0$, the state has $\langle \bm{S}^2 \rangle = 71.57$ and $\Delta E = 0.11t$, about one tenth of $W$.
Equation~\eqref{eq:eps_bound} then gives $\eta \leq 0.0272$, so that more than $97\%$ of the weight of the state lies in the sector $S = S_{\max}$.
This state is lower in energy by $1.33t$ than the HF-seeded state of the larger dimension $5.0\times 10^{9}$.
The HF-seeded and FM-seeded calculations use exactly the same measured bitstrings, so that changing only the reference occupation of iteration $0$ turns the low-spin state into a nearly fully polarized state that is much lower in energy.
No computed state has an energy below $E_{\mathrm{FM}}$, which is consistent with saturated ferromagnetism of the chain at $U = 5$.

If the ground state is saturated ferromagnetic, the HF-seeded calculations should also switch to nearly fully polarized states at larger dimensions, because the computed state approaches the ground state as the subspace approaches the full sector.
Subspaces of such dimensions, however, exceed the capacity of the HOKUSAI supercomputer at RIKEN used for these calculations.

\subsection{Energies and total spin of the $4\times 3$ lattice}
\label{sec:results:4x3}

\begin{figure}[tb]
  \centering
  \includegraphics[width=\columnwidth]{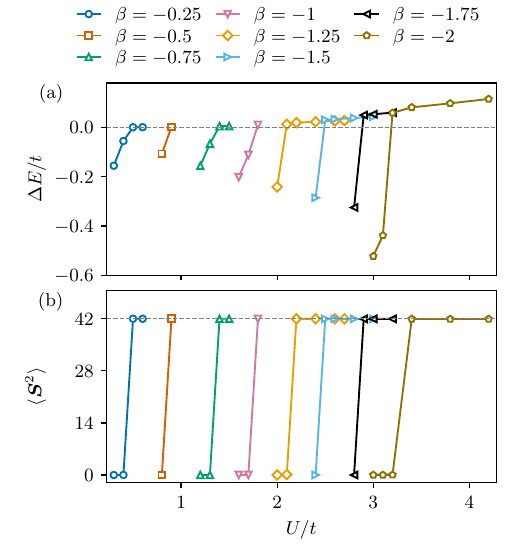}
  \caption{Energy and total spin of the FM-seeded calculations of the $4\times 3$ lattice as functions of $U$.
    (a) Energy measured from $E_{\mathrm{FM}}$.
    (b) $\langle \bm{S}^2 \rangle$, where the dashed line marks its maximum $S_{\max}(S_{\max}+1) = 42$.
    The symbols distinguish the values of $\beta$.}
  \label{fig:4x3}
\end{figure}

We next turn to the $4\times 3$ lattice with $M = 12$ unit cells, that is, $L = 24$ sites, where $N_e = 12$ and $S_{\max} = 6$.
As $\beta$ decreases from $-0.25$ to $-2$, the width of the lower band increases from $W = 0.20t$ to $1.63t$, so that the lattice moves away from the flat-band condition $\beta = 0$.
Figure~\ref{fig:4x3} shows $\Delta E$ and $\langle \bm{S}^2 \rangle$ of the FM-seeded calculations as functions of $U$ for $\beta = -0.25, -0.5, \ldots, -2$.
The subspace dimensions of the adopted states are between $5\times 10^{7}$ and $2\times 10^{8}$, which is less than $1\%$ of the full sector $S_z = 0$.

For every $\beta$, the computed state switches from a low-spin state to a nearly fully polarized state as $U$ increases.
The low-spin states have $\langle \bm{S}^2 \rangle \leq 0.14$, and the nearly fully polarized states have $\langle \bm{S}^2 \rangle \geq 41.92$.
Equation~\eqref{eq:eps_bound} then gives $\eta \leq 0.0067$ for the latter, so that more than $99.3\%$ of their weight lies in the sector $S = S_{\max}$.
The switching occurs within an interval of $0.1$ in $U$, except at $\beta = -2$, where the interval is between $U = 3.2$ and $3.4$.

Most of the low-spin states have $\Delta E < 0$, and the largest such $U$ gives the rigorous lower bounds $U_c > 0.4$, $0.8$, $1.3$, $1.7$, $2.0$, $2.4$, $2.8$, and $3.1$ for $\beta = -0.25, -0.5, \ldots, -2$ by the criterion of Sec.~\ref{sec:model:criterion}.
These bounds grow in proportion to the band width, and their ratio to $W$ lies between $1.9$ and $2.2$.

Above the switching, $\Delta E$ of the nearly fully polarized states increases from $0.0003t$ at $\beta = -0.25$ and $U = 0.5$ to $0.12t$ at $\beta = -2$ and $U = 4.2$, and is at most about $7\%$ of $W$.
Since $E_{\mathrm{FM}}$ does not depend on $U$, this increase reflects how well the subspace represents the fully polarized state.
The fully polarized state consists of the determinants that singly occupy the orbitals of the lower band, whereas the subspace is built from the RHF orbitals.
The RHF orbitals mix the lower and upper bands, because the mean-field potential $U\langle n_{i\downarrow}\rangle$ differs between the $a$ and $b$ sites.
Between the same two points, the weight of the fully polarized state outside the determinants that singly occupy the $N_e$ lowest RHF orbitals increases from $8\times 10^{-5}$ to $3\times 10^{-2}$, roughly in proportion to $\Delta E$.
No computed state above the switching has an energy below $E_{\mathrm{FM}}$, which is consistent with saturated ferromagnetism of the $4\times 3$ lattice at these values of $U$.

\subsection{Stability diagram}
\label{sec:results:diagram}

\begin{figure}[tb]
  \centering
  \includegraphics[width=\columnwidth]{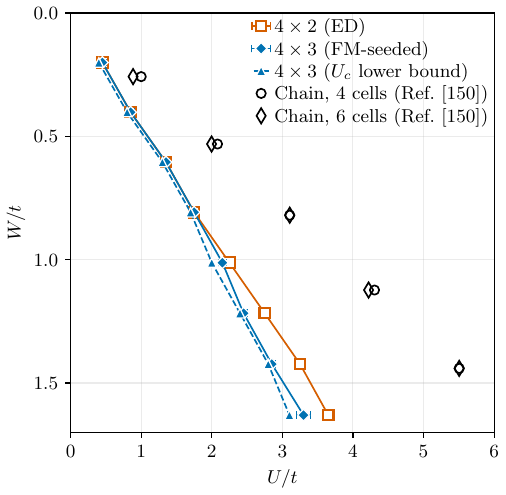}
  \caption{Stability of saturated ferromagnetism in the plane of $U$ and the band width $W$.
    The symbols with horizontal error bars show the intervals of $U$ that contain $U_c$ of the $4\times 2$ lattice obtained by exact diagonalization (ED) and the intervals in which the FM-seeded calculations of the $4\times 3$ lattice switch to the nearly fully polarized states.
    The triangles show the rigorous lower bounds on $U_c$ of the $4\times 3$ lattice obtained from states with $\Delta E < 0$.
    The open circles and diamonds show $U_c$ of the chains of 4 and 6 unit cells at quarter filling obtained by ED in Ref.~\cite{AritaAoki2000}, read from Fig.~16 of that reference.}
  \label{fig:stability}
\end{figure}

Figure~\ref{fig:stability} summarizes the stability of saturated ferromagnetism in the plane of $U$ and $W$.
As exact references, it includes the critical interaction $U_c$ of the chains of 4 and 6 unit cells obtained by exact diagonalization (ED) in Ref.~\cite{AritaAoki2000} and that of the $4\times 2$ lattice, which we obtain by ED.
On the chains, $U_c$ of 4 and 6 unit cells differ by less than $5\%$ for $W \geq 0.5t$, so that 4 unit cells along the chain already give $U_c$ close to that of the longer chain.
To approach the thermodynamic limit of the decorated square lattice, we therefore keep 4 unit cells along $x$ and examine how $U_c$ changes as the number of unit cells along $y$ increases from 2 to 3, that is, from the $4\times 2$ to the $4\times 3$ lattice.

On the $4\times 2$ lattice, the total spin of the ground state jumps at $U_c$ from $S = 0$ to $S_{\max}$ within the resolution $0.1$ of $U$, and the ground state above $U_c$ is the fully polarized multiplet with $E_0 = E_{\mathrm{FM}}$.
The critical interaction increases with $W$, and $U_c/W$ stays between $2.0$ and $2.3$ for $\beta \leq -0.5$.
As $W$ decreases toward the flat-band condition, $U_c$ decreases toward zero, in agreement with Tasaki's theorem that the fully polarized state is the ground state for every $U > 0$ when the lower band is flat.

The FM-seeded calculations of the $4\times 3$ lattice switch to the nearly fully polarized states within the same interval of $U$ as $U_c$ of the $4\times 2$ lattice for $\beta \geq -1$, and at values of $U$ smaller by $0.1t$ to $0.4t$ for $\beta \leq -1.25$.
The rigorous lower bounds give $U_c/W > 1.9$ on the $4\times 3$ lattice.
The critical interaction thus changes little from the $4\times 2$ to the $4\times 3$ lattice, which suggests that $U_c/W$ is also close to $2$ in the thermodynamic limit.

For the chains in the same range of $W$, the ratio is larger, $U_c/W \simeq 3.4$--$3.9$.
Saturated ferromagnetism is thus stabilized at a smaller $U/W$ on the decorated square lattice than on the chain.
This difference likely arises because each localized orbital of the lower band overlaps with more neighboring orbitals on the decorated square lattice than on the chain.
Because of $U$, electrons in overlapping orbitals favor parallel spins.
The number of overlapping neighbors of each orbital is roughly determined by $z$.

\section{Discussion}
\label{sec:discussion}

Whereas the main goal of previous applications of QSCI/SQD has been to obtain states with lower energies~\cite{Kanno2023,RobledoMoreno2025}, the present work uses the approximate wave function, which QSCI/SQD yields as an explicit vector in the subspace.
The total spin evaluated from this vector carries information on the computed state that the energy does not provide.
In the HF-seeded calculations of the chain, the energy decreases steadily with the subspace dimension, while $\langle \bm{S}^2 \rangle$ stays near zero and shows that the state remains far from the fully polarized multiplet.

For the nearly fully polarized states, $\langle \bm{S}^2 \rangle$ also shows quantitatively that they lie almost entirely in the sector $S = S_{\max}$, although its value in an approximate wave function is not quantized.
Because the eigenvalues of $\bm{S}^2$ are quantized, the largest one, $S_{\max}(S_{\max}+1)$, is separated from the others by at least $2S_{\max}$, so that a value of $\langle \bm{S}^2 \rangle$ close to $S_{\max}(S_{\max}+1)$ bounds the weight outside the sector of maximal spin through Eq.~\eqref{eq:eps_bound}.
This weight is below $3\%$ for the chain and below $0.7\%$ for the $4\times 3$ lattice.
This statement uses only the approximate wave function and requires no knowledge of the exact ground state.

Quantized quantities have been evaluated from approximate quantum states in other contexts.
Berry phases have been calculated from variational quantum states~\cite{Tamiya2021,MootzYao2026}, and the dimerized Hubbard chain studied in Ref.~\cite{MootzYao2026} provides an example of a quantized Berry phase that distinguishes topological phases.
Spin states of molecules have also been studied by QSCI/SQD~\cite{Liepuoniute2025}.
The present analysis combines the quantization of $\bm{S}^2$, which bounds the spin purity of the computed states, with the exact energy $E_{\mathrm{FM}}$ of the fully polarized state, which gives rigorous lower bounds on $U_c$.
The rigorous lower bounds rest on the fact that $E_{\mathrm{FM}}$ is known exactly for every $U$.
Any state with an energy below $E_{\mathrm{FM}}$ excludes saturated ferromagnetism, regardless of how the state is obtained.

Calculations on larger lattices with more qubits would allow a systematic extrapolation of $U_c/W$ to the thermodynamic limit.
Another direction for future work is to focus on other quantized quantities of the states obtained by QSCI/SQD, such as topological invariants.

\section{Summary}
\label{sec:summary}

We have studied saturated ferromagnetism in the Hubbard model on the sawtooth chain and the decorated square lattice with a dispersive lower band by QSCI/SQD with samples from IBM Quantum's Heron processors.
We characterized the computed states by their total spin and by their energy measured from the exact energy $E_{\mathrm{FM}}$ of the fully polarized state.

For the chain on 64 qubits, whose full sector $S_z = 0$ exceeds the $1.3\times 10^{12}$ determinants of one of the largest full configuration interaction calculations~\cite{Gao2024}, the FM-seeded calculation gives a state with more than $97\%$ of its weight in the sector $S = S_{\max}$ in a subspace of only $3\times 10^{-5}$ of the full sector.
The HF-seeded calculations give low-spin states at higher energies, although their energies decrease steadily with the subspace dimension.
For the $4\times 3$ decorated square lattice on 48 qubits, the FM-seeded calculations in subspaces of less than $1\%$ of the full sector switch from low-spin states to nearly fully polarized states with more than $99.3\%$ of their weight in the sector $S = S_{\max}$, and give a rigorous lower bound on the critical interaction $U_c$ for every $\beta$.
The switching occurs at $U \simeq 2W$, close to $U_c$ of the $4\times 2$ lattice obtained by exact diagonalization, and this weak dependence on the lattice size suggests $U_c \simeq 2W$ in the thermodynamic limit.
In these small subspaces the energies of the nearly fully polarized states remain above $E_{\mathrm{FM}}$ by up to about $W/10$, whereas $\langle \bm{S}^2 \rangle$ bounds their weight outside the sector $S = S_{\max}$ quantitatively from above.
The expectation value $\langle \bm{S}^2 \rangle$ thus characterizes the spin content of the computed states, while variational energies below $E_{\mathrm{FM}}$ provide rigorous lower bounds on $U_c$ of finite lattices.
This characterization requires only a very small fraction of the full configuration space, even for a lattice whose full configuration space exceeds that of one of the largest full configuration interaction calculations.

\begin{acknowledgments}
K.N. was supported by the UNIONTOOL Scholarship Foundation, the Uchida Energy Science Promotion Foundation, and MEXT Supporting Pioneering Research through AI for 1,000 Discovery challenges Program (SPReAD) Japan Grant Number JPMXP1726305043.
T.S. was supported by JSPS KAKENHI Grant No.~JP26K06972.
S.Y. was supported by JSPS KAKENHI Grant No.~JP21H04446. 
R.A. was supported by JSPS KAKENHI Grant Nos.~JP25H01246 and JP25H01252.
T.S. and S.Y. were additionally grateful for funding from JST COI-NEXT (Grant No. JPMJPF2221) and the Program for Promoting Research on the Supercomputer Fugaku (Grant No. MXP1020230411) by MEXT, Japan, as well as the New Energy and Industrial Technology Development Organization (NEDO) (project No.~JPNP20017). 
T.S. and S.Y. also acknowledge support from the COE research grant in computational science from Hyogo Prefecture and Kobe City through the Foundation for Computational Science.
Furthermore, we acknowledge support from the UTokyo Quantum Initiative and the RIKEN TRIP initiative (RIKEN Quantum, Advanced General Intelligence for Science Program, Many-body Electron Systems). 
The numerical calculations were carried out on Yukawa-21 at YITP in Kyoto University.
We acknowledge Center for Computational Materials Science, Institute for Materials Research, Tohoku University for the use of MASAMUNE-II (Project Nos.~202411-SCKXX-0002 and 202511-SCKXX-0002).
A part of the numerical simulations has been performed using the HOKUSAI supercomputer at RIKEN (Project ID RB250022).
The authors acknowledge the use of the AI-based writing tool OpenAI (GPT-6) and Claude (Opus 5.5) for language polishing. The authors take full responsibility for the content of this
\end{acknowledgments}

\appendix

\section{Qubit layout and device characteristics}
\label{app:hardware}

\begin{figure}[tb]
  \centering
  \includegraphics[width=\columnwidth]{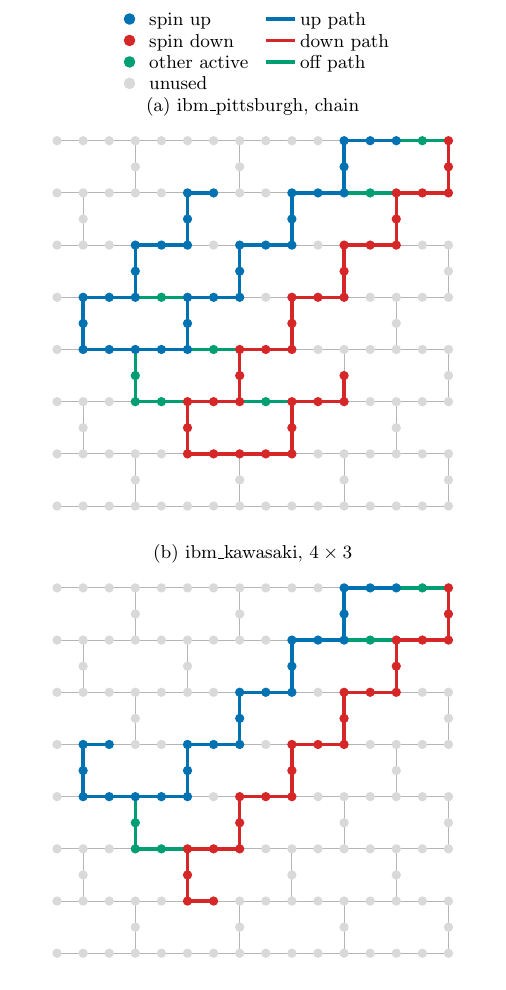}
  \caption{Qubit layouts of (a) the chain on \texttt{ibm\_pittsburgh} and (b) the $4\times 3$ lattice on \texttt{ibm\_kawasaki}, used for most of the circuits.
    Blue and red circles show the qubits of the spin-up and spin-down orbitals, green circles show the other active qubits, and gray circles show the unused qubits.
    Thick lines mark the couplings that carry a CZ gate in at least one transpiled circuit with the layout shown, and thin gray lines show the other couplings of the processor.
    Blue and red lines lie on the paths of the spin-up and spin-down qubits, which carry the orbital rotations and the same-spin pairs of $J$.
    Green lines mark the couplings outside these paths, which are used to implement the opposite-spin pairs of $J$.
    These pairs are the only two-qubit gates of the LUCJ circuit between qubits that are not adjacent in the initial layout, and the transpiler routes them through the green couplings.}
  \label{fig:layout}
\end{figure}

Figure~\ref{fig:layout} shows the qubit layouts of the chain and of the $4\times 3$ lattice.
The qubits of the spin-up and spin-down orbitals form two disjoint paths of coupled qubits.
The other active qubits connect the two paths, either as single ancilla qubits between them or as parts of the routes chosen by the transpiler.
Table~\ref{tab:device} lists the characteristics of the processors at the times of the experiments, where the two-qubit gate of these processors is the controlled-$Z$ (CZ) gate.

\begin{table}[tb]
  \caption{Characteristics of the IBM Quantum Heron processors used in this study: qubit relaxation time $T_1$, qubit dephasing time $T_2$, readout error, and CZ-gate error.
    For each quantity, the minimal, maximal, and average values over the qubits and CZ pairs used in all jobs are shown, taken from the calibration data at the time of each job.
    CZ-gate errors reported as $1$, which indicates that no calibrated value is available, are excluded.
    They amount to $8.9\%$ and $5.1\%$ of the CZ observations on \texttt{ibm\_kawasaki} and \texttt{ibm\_pittsburgh}.}
  \label{tab:device}
  \begin{ruledtabular}
    \begin{tabular}{lccc}
      Property & Min. & Max. & Avg. \\
      \hline
      \multicolumn{4}{l}{\texttt{ibm\_kawasaki}} \\
      $T_1$ [$\mu$s] & 1.3 & 618.9 & 285.1 \\
      $T_2$ [$\mu$s] & 2.2 & 976.6 & 193.9 \\
      Readout error & $1.10\times 10^{-3}$ & $4.34\times 10^{-1}$ & $2.17\times 10^{-2}$ \\
      CZ-gate error & $4.75\times 10^{-4}$ & $1.94\times 10^{-1}$ & $9.68\times 10^{-3}$ \\
      \hline
      \multicolumn{4}{l}{\texttt{ibm\_pittsburgh}} \\
      $T_1$ [$\mu$s] & 6.6 & 474.5 & 269.3 \\
      $T_2$ [$\mu$s] & 8.6 & 633.1 & 306.1 \\
      Readout error & $1.46\times 10^{-3}$ & $1.02\times 10^{-1}$ & $1.58\times 10^{-2}$ \\
      CZ-gate error & $8.19\times 10^{-4}$ & $2.07\times 10^{-1}$ & $8.43\times 10^{-3}$ \\
    \end{tabular}
  \end{ruledtabular}
\end{table}

\section{Calculations without initial recovery}
\label{app:raw}

This appendix compares the FM-seeded calculations of the $4\times 3$ lattice in Sec.~\ref{sec:results:4x3} with calculations in which iteration $0$ uses no configuration recovery.
In these calculations without initial recovery, the configurations measured in iteration $0$ are only post-selected and subsampled, and the iterations $i \geq 1$ proceed as described in Sec.~\ref{sec:methods:recovery}.
These calculations use between five and ten iterations, whereas those of Sec.~\ref{sec:results} use five.
The occupations obtained in iteration $0$ are close to the HF reference at most of the computed parameters, so that the calculations without initial recovery proceed almost as HF-seeded calculations.
The samples are taken from the circuit generated at $U = 1.4$ for $U = 1.7$ and $2.5$ at $\beta = -0.5$, at $U = 3.2$ for $U = 3.75$ and $5$ at $\beta = -1$, at $U = 4.5$ for $U = 5$ at $\beta = -1.25$, and at $U = 9.3$ for $U \geq 10$ at $\beta = -2$, and otherwise from the circuit generated at the same value of $U$.

\begin{figure}[tb]
  \centering
  \includegraphics[width=\columnwidth]{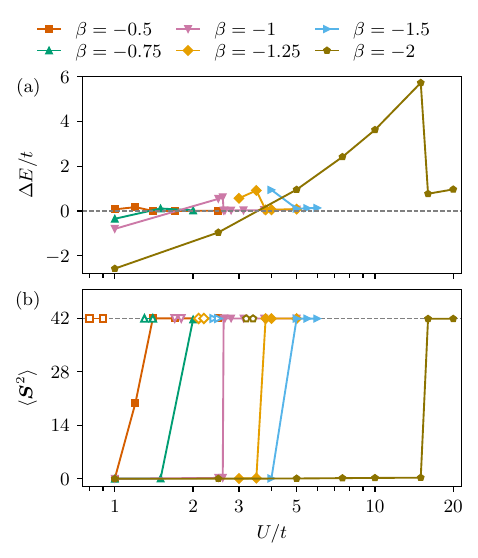}
  \caption{Energy and total spin of the calculations without initial recovery of the $4\times 3$ lattice as functions of $U$.
    (a) Energy measured from $E_{\mathrm{FM}}$.
    (b) $\langle \bm{S}^2 \rangle$, where the dashed line marks its maximum $S_{\max}(S_{\max}+1) = 42$.
    The symbols distinguish the values of $\beta$ as in Fig.~\ref{fig:4x3}.
    The pairs of open symbols in (b) mark the intervals in which the FM-seeded calculations switch to the nearly fully polarized states, found in Fig.~\ref{fig:4x3}(b).
    The horizontal axis is logarithmic.}
  \label{fig:raw}
\end{figure}

Figure~\ref{fig:raw} shows $\Delta E$ and $\langle \bm{S}^2 \rangle$ of these calculations as functions of $U$.
The calculations without initial recovery also switch from low-spin states to nearly fully polarized states as $U$ increases, but at a larger $U$ than the FM-seeded calculations for every $\beta$.
The difference is largest at $\beta = -2$, where the switching occurs near $U = 15$ instead of $U \simeq 3.3$.
Between the two switching points, the low-spin states lie above $E_{\mathrm{FM}}$, whereas the nearly fully polarized states of the FM-seeded calculations lie close to $E_{\mathrm{FM}}$.
The lower bounds on $U_c$ obtained from the states with $\Delta E < 0$ are consistent with those of the FM-seeded calculations, which are sharper.

\clearpage
\bibliography{refs}

\end{document}